\documentstyle[11pt,aaspp4]{article}

\begin{document}

\title{PKS 0116+082: An Optically Variable Compact Steep-Spectrum Source in an NLRG}

\author{M.H. Cohen, R.C. Vermeulen\altaffilmark{1}, P.M. Ogle}
\affil{Palomar Observatory, California Institute of Technology,
Pasadena, CA 91125}
\authoremail{mhc@astro.caltech.edu}

\author{H.D. Tran\altaffilmark{2}} 
\affil{UCO/Lick Observatory, University of California, Santa Cruz, CA 95064}
\and
\author{R.W. Goodrich}
\affil{W.M. Keck Observatory, 65-1120 Mamalahoa Highway, Kamuela, HI 96743} 

\altaffiltext{1}{Current address: NFRA, Postbus 2, NL-7990, AA Dwingeloo, Netherlands}

\altaffiltext{2}{Current address: Lawrence Livermore National Laboratory,
P.O.Box 808, L-413, Livermore, CA 94551}

\begin{abstract}

   Polarimetry of the narrow-line radio galaxy PKS 0116+082 at the W.M.
Keck telescope shows that it has high and variable optical
polarization, presumably due to synchrotron radiation.  It is
not a BL Lac object because it has strong narrow lines, and it is not an OVV
quasar because it has no broad lines and the extended galaxy is
prominent.  VLA and VLBA images show that it is a compact
steep-spectrum radio source with most of the emission coming from a
region less than 100 milli-arcsec in size.  Of the 25 compact
steep-spectrum or gigahertz peaked-spectrum sources measured
polarimetrically, four have high optical polarization. One of these has
been observed only once but in the other three the polarization is
variable.  This gives an intriguing hint that variability may be a
general property of these objects.

\end{abstract}


\section{Introduction}

   We have been measuring the polarization of radio galaxies with the
W. M. Keck telescope, to study the alignment effect, the composition
and disposition of material around the nucleus, the redshift dependence
of the polarization, and the possible unification of narrow- and
broad-line radio galaxies (NLRG and BLRG). For a general review of these
topics see \cite{m93}.  As part of our study we observed the galaxy
associated with the radio source PKS 0116+082 (J0119+0829).  This
source was identified with a distant galaxy by \cite{s67} with $z =
0.594$ and $m_v = 20.7$, and spectroscopically is an NLRG.  The
optical radiation shows high polarization, as measured by \cite{t92}
who found $p = 5.4 \% \pm 0.6\%$ at position angle $\theta = 75\arcdeg
\pm 3\arcdeg$, in the $V'$ band (approximately 4885 -- 6235 \AA).

  The radio source has not been well-studied; e.g., it has no VLA or
VLBI images in the literature.  However, it does display strong
interplanetary scintillations, and therefore must be compact on a scale
of one arcsecond. \cite{h69} showed that most of the source must lie
within $0.05''$ at 408 MHz, and \cite{r74} showed that at least 0.65 of
the flux comes from within $0.45''$ at 81.5 MHz.  This object clearly
is different from most other high-redshift radio galaxies (HZRG), in
which most of the radio flux comes from extended radio lobes that have
conspicuous hot spots near their outer edges (FRII morphology).

  PKS 0116+082 may have flux variations of order 20\% at centimeter
wavelengths (\cite{gc91}; \cite{gr95}), but there is no evidence for
large or rapid fluctuations. The spectrum is straight from 100 to 10,000
Mhz, with an index $\alpha = -0.52$ ($F_\nu \sim \nu^\alpha$;
\cite{k81}).  There is no suggestion of flattening at high frequencies,
although there may be a steepening below 100 MHz.  This index puts the
source on the boundary between ``steep" and ``flat" spectrum objects.
It appears to belong to the category generally designated as
``compact steep-spectrum" (CSS; see e.g., \cite{o91}).  In these objects
nearly all the radio emission arises within the host galaxy,
although the radio image typically resembles that of the larger
sources. A small fraction of CSS objects has complex or
``two-dimensional" structure (\cite{s95}), as does PKS 0116+082 (see
below).  It is thought that this compact complex structure arises from
interactions between a radio-emitting relativistic jet and a dense
interstellar medium near the nucleus of the galaxy (e.g., \cite{w91}).

   Although there are no published VLA images of PKS 0116+082, this
object indeed has been observed with the VLA, on several occasions. W.
Coles and R. Grall observed it in 1990 May and June, and have kindly
shared their original data with us. We show the first VLA\footnote{The
VLA and the VLBA are facilities of the NRAO which is operated by AUI
under a cooperative agreement with NSF.} images below.  In addition, A.
Beasley was able to obtain a VLBA$^3$ snapshot for us and that is
presented below also. These results show that the radio designation CSS
is correct.

   Our optical observations, described below, show that the optical
polarization is strongly variable. This seems surprising because PKS
0116+082 is not BL Lac-like with very weak lines, nor is it quasar-like
with broad emission lines and little or no starlight.  However, little
is known about the polarization of CSS objects, and the similar
gigahertz peaked-spectrum (GPS) objects. In Section 4 we discuss the
relation of PKS 0116+082 to other CSS/GPS objects.

   In this paper we assume $\rm{H_o} = 65 ~km ~sec^{-1} ~Mpc^{-1}$ and
$\rm{q_o} = 0.5$.

\section{Optical Observations}

  a) {\bf Observations}

   We measured the polarization of PKS 0116+082 using the
Low-Resolution Imaging Spectroscope (LRIS, \cite{o95}) on the 10-m
telescope at the W.M. Keck Observatory. The instrument contains a
beamsplitter giving simultaneous measurements in orthogonal linear
polarizations, and uses a half-wave plate set successively to four
position angles, 0$\arcdeg$, 45$\arcdeg$, 22.5$\arcdeg$, and
67.5$\arcdeg$ as described by \cite{g91}. Further aspects of the
polarimeter are given by \cite{g95}, and in Appendix A. Table 1 gives
the journal of observations.  For spectroscopy we used a 300 g/mm
grating blazed at 5000\AA, which gives a resolution of 10\AA ~with the
1$''$ slit and 15\AA ~with the 1.5$''$ ~slit.  Flux calibration was
done with various standard stars: GD 248, G 193--74, G 191B2B, Feige 34,
and Feige 110.  Assorted clouds and highly variable seeing made most of
the nights far from photometric.

  For epochs 4a and 4b (Table 1) we replaced the grating in LRIS with a
plane mirror and made imaging polarimetric observations through a
Johnson $B$ filter.  Unpolarized standard stars (e.g., \cite{s92}) could not
be used for primary calibration because they are too bright, and we
tested the imaging polarimetry with stars 137 and 139 in the field
Landolt 95 (\cite{l92}). These stars have $B$ magnitudes 15.86 and 13.12
respectively.  In high and variable seeing we measured, through the
polarimeter, a flux ratio of $2.78 \pm 0.02$ mag, which is in good
agreement with Landolt's values.  We observed this field both with and
without the UV calibration polarizer, and the polarization of the stars
was measured by summing the appropriate fluxes in a square 3$''$ on
edge and then calculating the Stokes parameters.  With the polarizer
stars 137 and 139 gave $p = 99.5\% \pm 0.2\%$ and $99.9\% \pm 0.1\%$,
respectively.  Without the polarizer they gave $p = 0.0\% \pm 0.16\%$
and $0.2\% \pm 0.04\%$, respectively.  The errors are calculated from
photon statistics. The galactic latitude of Landolt 95 is $-37\arcdeg$,
sufficiently high that it is unlikely that interstellar polarization
had any appreciable effect on these measurements.  While there is no
guarantee that the Landolt standards have low polarization, it is
unlikely that they would be significantly polarized at a level and
angle which would cancel any instrumental polarization.  It is more
likely that both the instrumental polarization and the polarization of
the Landolt stars are near zero.  The system is working properly at the
extremes of 0\% and 100\% polarization.

   The seeing was poor and variable during the 1995 December run,
particularly for the observations of Landolt 95.  Nonetheless, the
polarimetry gives good results. This is because a dual-beam polarimeter
is robust against seeing and transparency variations, provided an
appropriate reduction algorithm is used.  See Appendix A for a
discussion of this point.

  b) {\bf Spectropolarimetry Results}

   We have evidence that the polarization in PKS 0116+082 is variable
(Section 2d) and hence do not combine data from different epochs.
In Figure 1 we show the flux and polarization data for epoch 5a, which
has the highest signal-to-noise ratio (S/N). The spectrum in Figure 1a is
that of a typical NLRG.  \cite{g94} studied the optical spectra of a
sample of 19 CSS sources, with the view of comparing them with other
categories of objects and testing the idea that the small size is
connected with an interaction between the interstellar medium and a
relativistic jet.  In all but one of Gelderman and Whittle's objects
[OIII]5007/H$\beta >> 1$ as in Figure 1a, but in those objects where both
[OII] and [OIII] are seen only one of ten has [OII]3727/[OIII]$5007>1$.
PKS 0116+082 appears to have somewhat lower ionization than the typical
CSS source.

   In Figure 1b,c,d the fluxes were binned by 4 pixels to improve the
S/N, and then the polarization parameters were calculated. Figure 1b shows
the position angle $\theta$ which is independent of $\lambda$ from 4200
to 7600 \AA\, and has a mean value $\langle\theta\rangle = 102.4
\arcdeg \pm 0.6 \arcdeg$.  Figure 1c shows the ``rotated Stokes parameter"
$q'$ which is an estimate of the true linear polarization $p$ obtained by
rotating $q$ and $u$ by --204.8$\arcdeg$ (see Appendix A).  The polarized
flux (Fig 1d, the product of Figs 1a and 1c) is smooth and rises to the
blue with a spectral index $\alpha \approx -1.5$. Note that the narrow
lines do not appear in the polarized flux, indicating that they are
unpolarized.  This can also be seen in Figure 1c where $p$ drops to a low
level in the [OII] line. The formal polarization of the [OII] line is
$p([\rm{OII}]) = 0.4\% \pm 0.6\%$.

   In Figure 1a several prominent galactic absorption lines are seen.  We
used a Keck spectrum of NGC 6702 as a template and found that the
galaxy fraction is $29\% \pm 2\%$ at $\lambda_{\rm rest} = 4600$\AA,
and about 12\% near $\lambda_{\rm rest} = 3727$\AA.  The AGN component
(total -- galaxy) has $p = 4.9\% / 0.71 = 6.9\%$ at $\lambda_{\rm rest}
= 4600$\AA\ and $p = 7.3\% / 0.88 = 8.3\%$ at $\lambda_{\rm rest} =
3727$\AA. This suggests that $p$(AGN) rises to the blue, but the errors
inherent in this calculation are fairly large, and the increase may not
be real. In any event, the apparent increase in $p$ to the blue, 
seen in Fig. 1c, is largely due to the decrease in galaxy dilution.

  The spectra of total and polarized flux show no significant broad
H$\beta$ or Mg II emission, although the galaxy--subtracted spectrum
(the AGN spectrum) gives a weak hint of broad H$\beta$.  The best way
to test for broad lines would be to look for H$\alpha$, at
1.05 microns.

 c) {\bf Imaging Polarimetry Results} 

   Figure 2 shows the $B$ band polarization image of PKS 0116+082, made
by combining the data from epochs 4a and 4b. The data are binned 2x2 so
that each pixel is 0.43$''$ square. Only those polarization vectors
with $\rm{S/N} > 2.0$ are shown. The strongest of the faint objects to
the north has a formal debiased polarization $p = 0.8 \pm 0.9\%$, and
the NW object has $p = 0.1\% \pm 0.1\%$. These low values are
consistent with these two objects being unpolarized, and give
confidence that the system is working properly.

  Note in Figure 2 that the polarization is uniform across the bright
central part of the galaxy, as expected for a point source broadened by
seeing. The PSF's of PKS 0116+082 and the NW object are approximately
equal, with FWHM=1.1$''$. This galaxy does not show the polarized
extensions which are common in other HZRG. (See e.g.,
Tadhunter et al. 1992; \cite{d94}; \cite{ja95}.) The central square
2.4$''$ of PKS 0116+082 has $p = 15.9\% \pm 0.2\%$ and $\theta = 93.2
\arcdeg \pm 0.2 \arcdeg$.  Both $p$ and $\theta$ are substantially
different from their 1991 values, as reported by \cite{t92}. This
source is clearly variable, and in the next section we present all our
data and discuss the variability.

 d) {\bf Polarization Variability}

   In Table 2 we list our polarization measurements, and also the
values given by \cite{t92} from their observations in
1991.  The values for the imaging data were obtained by summing the
fluxes in a box which simulates the spectral extraction window, and
then calculating the Stokes parameters. The values for the
spectroscopic data were made by similarly summing in wavelength to
simulate $B$ and $V$ bands.  Clearly these simulations are not exact, but
tests of small changes in the windows produced only small changes in
$p$ and $\theta$, less than the stated errors. The observations for epoch 2
were taken through clouds and the data below about 4600\AA\ are not
reliable.  Our spectroscopic observations in 1994 were taken with a
1$''$ slit and those in 1995 with a 1.5$''$ slit.  Dilution of the
polarization by starlight or by any extended unpolarized
component would cause $p$ to be smaller in 1995 than in 1994, contrary
to what we observed. Further, slit width combined with dilution cannot
change $\theta$ because Figure 2 shows that $\theta$ is essentially
constant across the central region.

   Table 2 shows that $p$ had about the same values in 1991 and 1994
and that there is little significant difference between the $B$ and $V$
band values. However, the temporal changes in $\theta$ are highly
significant.  In 1995 the values of $p$ and $\theta$ from
measurement sets one hour apart (a and b) show changes; but these, we
suspect, show that the errors (calculated from the photon statistics)
are underestimated. However, the decrease in $p$ from epoch 4 to epoch
5, one day later, is surely significant; and the change in $\theta$ is
also significant, but at a lower level.

   The high value of $p$ from the imaging data, especially when
compared with the spectral data for the following night, calls for a
rigorous test of the system and the reduction procedure before the
results can be fully accepted.  A strong test is to compare the imaging
data with spectral data taken on other occasions.  We observed six
objects in imaging mode on 1995 December 16.  Four of the six have
values of $(p,\theta)$ which agree with our spectropolarimetry to
within $(0.4\%, 2.3\arcdeg)$.  One has $\Delta p \approx 0.3\%$ and
$\Delta\theta \approx 6\arcdeg$. The sixth, PKS 0116+082, is much more
discordant.  We conclude that the data are trustworthy, and that the
polarization of PKS 0116+082 changed strongly in one day.

   Additional evidence of variability comes from the equivalent
widths.  Table 2 lists the rest-frame equivalent widths W$_\lambda$ of
the [OII] $\lambda3727$ lines, in Angstroms. The 1995 values are
significantly smaller than those from 1994. The differing slit widths
in principle could affect W$_\lambda$ because the starlight and [OII]
emission could be extended but the UV continuum might be point-like.
However we think this effect is small because seeing could similarly
cause changes in W$_\lambda$, but no changes are seen between epochs 5a
and 5b.  It is clear that W$_\lambda$([OII]) decreased from 1994 to
1995, while $p$ increased. 

   From the polarization and the [OII] equivalent width we calculate
$F^\prime_c$, the AGN continuum flux relative to the integrated [OII]
flux (times 10,000), and $p_c$, the polarization of the AGN continuum
at 3727\AA.  In doing this we assume that the [OII] flux is
constant and that the galaxy component is constant and unpolarized.
The appropriate formulae are

	 $$p_c = {p \over {1-g}};
   ~~~~~~~F^\prime_c \sim {F_c \over \rm[OII]} = {{1-g} \over \rm W}$$

\noindent where $g$ is the galaxy fraction at 3727\AA\, and W is the
equivalent width of [OII]. For earlier epochs when the galaxy fraction
could not be accurately measured, we replace $g$ with
$g_5(\rm{W/W_5})$, where $g_5$ and $\rm{W_5}$ are measured at epoch 5a.
$F^\prime_c$ and $p_c$ are given in Table 2, columns 7 and 8. From 1994
to 1995 the AGN flux increased by about 30\%, and the polarization from
about 6\% to 9\%.

   Note that the implied equivalent widths of the emission lines are
well within the normal observed range for AGN (e.g., \cite{d88}); there
is no evidence for a deficit of ionizing photons from the line
strengths.  This is especially true given the steepness of the
galaxy-corrected continuum, $\alpha = 0.72$.

\section{Radio Observations}

    PKS 0116+082 was observed for an unrelated project at the VLA by
W. Coles and R. Grall in 1990 May and June at three frequencies, 1.7,
5.0, and 8.5 GHz, in A configuration with 50 MHz bandwidth, in single
scans of duration 70, 90, and 100 sec at the three frequencies,
respectively.  Coles and Grall have kindly shared their original data
with us, and we produced images from them.  The brief snapshots have
poor dynamic range, but the resulting images are adequate to obtain a
reasonable idea of the shape of the source and its orientation. Two
8.5 GHz images are shown in Figure 3. The ``tapered" map (giving
maximum sensitivity to extended emission) in Figure 3a shows a weak
component to the east extending about $2''$. The ``uniformly-weighted"
map in Fig. 3b (giving maximum angular resolution at the expense of
dynamic range) shows a faint extension to the west and, in
addition, this image shows a more prominent extension to the
north-west, close to the center. Model-fitting two circular Gaussians
to the inner extension gives component strengths of 0.52 and 0.14 Jy.
Their sizes are not well-constrained but they are separated by
$\sim$225 milli-arcsec (mas) ($\sim$1 beamwidth) in position angle
$-47\arcdeg$.  The 5.0 GHz observations also show both an eastern and a
(north)western extension, but the inner double is not well resolved.
The observations at 1.7 GHz clearly show only the eastern extension.
These results show that PKS 0116+082 is compact at centimeter
wavelengths, with nearly all the flux arising within 1 arcsec.  This
confirms the results from interplanetary scintillations.

   PKS 0116+082 was observed at the VLBA by A. Beasley at 1.7 GHz on
1996 March 14. The observation consisted of one 5-minute
snapshot, and the resulting image, shown in Fig. 4a, has only a modest
dynamic range. The source was then included by RCV in an unrelated
multi-snapshot VLBA survey at 5.0 GHz, observed on 1996 August 22. The
image, which is essentially thermal noise limited, is shown in Fig. 4b.

    The images show an elongated curved structure, with an overall
length of about 75 mas, or 0.5 kpc. The nucleus is probably in the
northernmost, most compact component. This feature has an inverted
spectrum and is not the strongest one even at 5.0 GHz, presumably due
at least in part to synchrotron self-absorption.  The brightest
component at both frequencies occurs south of the gap, and
seems to mark the onset of a region of strong curvature, in which the
jet definitely becomes resolved transversely. Afterwards, the jet
points almost directly at the knot seen with the VLA (Fig. 3b) at a
projected distance of 225 mas.

a) {\bf A New Radio Position}

   The 8.5 GHz VLA observations allow us to calculate a new radio
position for the source, given in Table 3.  This position was obtained
by model-fitting to the pre-selfcalibrated VLA data, which were
calibrated with the astrometric calibrator J0149+059 (\cite{j95}). 
Model-fitting errors are about 0.02$''$ and the phase transfer
errors are probably of similar magnitude; we estimate the positional
errors in both RA and Dec to be about 0.035$''$.  Table 3 also contains
the \cite{s67} value for the optical position, precessed to
J2000 coordinates. The radio and optical positions agree to within the
stated error in the optical position.

\section{Discussion}

a) {\bf Classification of PKS 0116+082}

   PKS 0116+082 is in an NLRG displaying rapid optical polarization
variability. The two classes of objects with such variability are BL
Lac objects and optical violently variable (OVV) quasars, but PKS
0116+082 cannot be placed in either one. The definitions of these
classes are not tight because observers emphasize different
characteristics, which themselves generally have a broad distribution.
Furthermore, some objects have ambiguous classifications because their
variability carries them from one class to another.  We shall borrow
from the discussion of \cite{j93} who, in a discussion of X-ray
selected BL Lac objects, reviewed the history of the definitions of
some of the AGN classes. A BL Lacertae object, according to Januzzi et
al., is an object with (1) strong and rapid variability, (2) a
``featureless" optical spectrum (W$_\lambda < 5$\AA), and (3) high
linear polarization.  A highly-polarized quasar (HPQ) is a quasar with
optical polarization $p > 3\%$.  An OVV quasar displays rapid and large
changes in flux density. It appears that all OVV quasars are also
HPQ's, but the converse is not true. A blazar is any object which is
violently variable in flux and has high and variable optical
polarization.  Blazar is a general term which includes BL Lacs and OVVs
even though these typically have very different luminosities.

   The important feature of PKS 0116+082 is that it has strongly
variable optical polarization. However it does not appear to have
strong radio variability, so it is not a blazar. It is not a BL Lac
because it has strong narrow lines and it is not a quasar because it is
clearly extended, and furthermore any broad lines are very weak.  It
cannot be classified in the conventional way.

b) {\bf Origin of the Polarization}

   The optical light from AGN can be polarized by selective absorption
in aligned dust grains, by scattering from electrons or dust, and by
the synchrotron effect. Dust absorption appears to be ruled out in PKS
0116+082 because the high polarization (16\%) is extreme for dust
absorption, and because polarization due to this mechanism cannot vary
on short time scales. 
Similarly, although polarization caused by scattering off
dust or electrons can easily reach the high values seen in PKS
0116+082, it cannot easily have a position angle which varies on short
time scales. We therefore propose that the polarization in PKS 0116+082
is due to synchrotron radiation.  The lack of polarization in the
emission lines (Section 2b) is consistent with this interpretation.

   The optical synchrotron source must lie well inside the radio
synchrotron source, because it requires a combination of higher density
and stronger magnetic field. The variation in one day implies that it
is not larger than about $3 \times 10^{15}$ cm in extent. This estimate
is too small if the optically emitting synchrotron material itself has
an outward bulk relativistic motion.

c) {\bf Relation to Other CSS/GPS Sources}

   CSS radio sources are compact and have a steep spectrum. The GPS
radio sources are similar except that their spectrum has a peak around
1 GHz. The peak is presumably due to synchrotron self-absorption, and
the only distinction between the two classes seems to be that the GPS
sources have higher optical depth to synchrotron radiation.  We can
consider the CSS and GPS sources together.

   About 150 CSS/GPS sources are known (\cite{d90}, \cite{o91}), and we
have found optical polarization measurements for 25 of them in the
literature (\cite{i90}, \cite{i91}, \cite{t94}, \cite{x95}, and our
unpublished Keck data).  8/25 are definitely polarized ($p \ge
3\sigma$) and 4/8 are highly polarized ($p \ge 3\%$). Most of the
measured objects are quasars (22/25) and the remaining 3 are NLRG. Of
the highly-polarized objects, 2 are quasars and 2 are NLRG. The
fraction of HPQ among the quasars (2/22) appears to be smaller than the
HPQ fraction (0.3) found by Impey and Tapia, but most of the quasars in
our list were only observed once and the fraction would surely increase
if a serious measurement campaign were made.

  The two HPQs are 3C 216 and CTA 102. 3C 216 (z=0.67) originally
was called a BL Lac object but then was labelled a quasar by \cite{s80} 
because it showed substantial [OII]$\lambda$3727 and,
perhaps, broad MgII $\lambda$2800.  \cite{l96}, with a better
spectrum, list the equivalent widths of [OII]$\lambda3727$ and
[OIII]$\lambda5007$ as 15.5 and 13.0 \AA, respectively, which are
larger than the usual BL Lac limit, 5\AA.  The FWHM of MgII and
H$\beta$ are both 1800 km sec$^{-1}$ which is intermediate between
``narrow" and ``broad".  At radio wavelengths 3C 216 is a CSS object
containing superluminal components (\cite{b88}) but, like PKS
0116+082, it is not a rapid variable and would not be called a blazar.
Its optical polarization is variable (\cite{k76}), with maximum
21\%.

  CTA 102 (z=1.037) also has variable optical polarization (\cite{i90})
with maximum 10.9\%. At centimeter and meter wavelengths its flux has
slow variations. The existence of superluminal motion in CTA 102 has
been debated (\cite{w89}, \cite{r96}), but there is no doubt that CTA
102 contains a powerful radio jet. It is also a $\gamma$--ray source
(\cite{t95}).

   The two highly-polarized NLRG are PKS 0116+082 and PKS 1934--638.
Tadhunter, Shaw \&\ Morganti (1994) showed that PKS 1934--638 ($z =
0.183)$ has strong optical polarization in the $B$ band, 4\% in a small
aperture and perhaps as much as 7\% for the featureless continuum
alone.  It has only been observed once, as far as we are aware, so that
its variability is unknown.  Fosbury et al. (1987) discussed its
spectrum; it has strong narrow lines which on analysis show that the
electron density is higher than usual for NLRG.  The radio morphology
is double, with an exceptionally large ratio of separation to size.
There is no superluminal motion (\cite{t89}).

   PKS 0116+082 and PKS 1934--638 have similar optical spectra and both
are highly-polarized.  They are compact at radio wavelengths but the
detailed small-scale structure is different. PKS 0116+082 is complex,
whereas PKS 1934--638 is a stationary double with a large ratio of
separation to size; these differences may reflect the density and
complexity of the ISM near the galactic nucleus.  All three of the
well-measured highly polarized CSS/GPS objects show strongly variable
polarization. It will be interesting to measure the fourth object, PKS
1934--638, to see if it, too, is variable.

d) {\bf Nature of PKS 0116+082}

   We first ask if the unusual nature of PKS 0116+082 can result from
an accidental close alignment of two objects; one a CSS source in
an NLRG, and the other an optically variable object either in front of
or behind the NLRG. The spectrum contains no hint of a second redshift
system, so the variable object would have to be lineless; ie. a BL Lac
object.  We think this is highly unlikely because both CSS/GPS and BL
Lac objects are rare. The similarity of PKS 0116+082 and PKS 1934--638
argues for a common mechanism, which makes the probability of chance 
alignments between the two rare classes vanishingly small.

   The synchrotron origin for the polarization implies that the nucleus
contains high-energy electrons in a magnetic field. The rapid
variations mean that we probably are getting a direct view of this
nuclear source rather than seeing it by reflection or through a cloud
which would quench the rapid variations.  On the other hand, many HZRG
are believed to contain an opaque torus that confines the nuclear
optical and UV light to oppositely directed ``cones" of emission (e.g.,
IRAS 09104+4109, \cite{h93}).  If PKS 0116+082 has such a torus then
either we are looking close to the axis, or the torus has a ``hole"
through which we have a view of the central continuum source. In the
former case we should also see any broad-line region (BLR) which exists near
the nucleus, but we see little evidence for broad lines.  In the latter
case the hole could probably be arranged to exclude a BLR, but this
seems contrived.  We note that spectra of PKS 1934--638 (\cite{f87})
also give no evidence of broad H$\alpha$ or H$\beta$.  We suggest that
the lack of observed broad lines in PKS 0116+082 is best interpreted as
meaning that in fact there is no BLR, or at least that it is unusually
weak.  In this case it does not fit into the quasar/BLRG/NLRG
unification scheme (e.g., \cite{b89}).

  We assumed above that the radio nucleus of PKS 0116+082 is the
northern compact component, and described the structure as ``core-jet"
with a pronounced curvature beginning about 60 mas (400 pc, projected)
from the nucleus.  The one-sided nature of the jet is usually regarded
as a sign of relativistically boosted synchrotron radiation, and this
leads to limits on the Lorentz factor of the flow, $\gamma$, and the
angle to the line of sight, $\phi$. For example, if the front-to-back
ratio is 20 or more, then $\gamma > 1.3$ and $\phi < 50\arcdeg$,
approximately. The strong curvature is also a sign of an end-on view,
in which a modest curvature can be strongly amplified by projection.
These factors argue that the radio jet is not close to the plane of the
sky, say $\phi < 50\arcdeg$.  This reinforces the possibility that we
are looking inside any central torus, and should see a broad
emission-line region if it were there.

  The three CSS/GPS objects with variable polarization are united by
radio morphology and spectrum, but optically they include both quasars
and narrow-line radio galaxies. Furthermore, in the case of PKS
0116+082, it appears doubtful that the RG/quasar unification-by-aspect
is at work.  Since the polarized CSS/GPS objects transcend both
galaxies and quasars, and because their number is so small, it does not
seem useful to try to define a new subclass for them. Further
polarimetric observations would be useful in this regard.

\section{Conclusions}

1) PKS 0116+082 shows rapid variations in optical polarization,
including a change from 16\% to 9\% in one day. The position angle has
varied from $10\arcdeg$ to $103\arcdeg$ over a one-year period. Although
our observations were typically not in photometric conditions, we were
able to scale the polarized flux by assuming that the
[OII]$\lambda3727$ flux and the host galaxy flux are constant. From
1994 to 1995 the AGN continuum flux increased by about 30\%, while the
AGN polarization increased from about 6\% to 9\%.

2) PKS 0116+082 is a CSS radio source in an NLRG. The radio structure
is compact and complex. We show the first VLA and VLBI images of this
object.

3) The radio flux appears to have only modest variations. PKS 0116+082
is not a BL Lac object because it has strong narrow lines and it is not
a HPQ because the host galaxy is prominent and there are no readily
visible broad lines.

4) Of the approximately 150 known CSS/GPS sources, 25 have been
observed polarimetrically. Four are known to be highly
polarized at optical wavelengths. Three of the four have variable
polarization and the fourth has been observed only once.  

5) Our current data give little evidence for broad H$\beta$ or Mg II
emission in the total or polarized flux from PKS 0116+082, although
optical synchrotron radiation is seen. The most likely explanation for
this is that any broad-emission-line region is exceptionally weak.

\acknowledgments

   We are grateful to W. Coles and R. Grall for giving us their raw VLA
data, and to A. Beasley for observing PKS 0116+082 for us on the VLBA.
We had valuable discussions with P. Barthel, who pointed out the
relevance of 3C 216; with R. Fosbury, who pointed out the relevance of
PKS 1934--638; and with I. Browne, C. O'Dea and T. Pearson. We thank the
anonymous referee for useful suggestions.  The W.M. Keck Observatory is
operated as a scientific partnership between the California Institute
of Technology and the University of California, and was made possible by
the generous financial support of the W.M. Keck Foundation.  This
research has made use of the NASA/IPAC Extragalactic Database (NED).
This work was partly supported by NSF grant AST91-21889.

\appendix
\bigskip\bigskip
\centerline{\bf APPENDIX}
\section{The LRIS Polarimeter}

   The LRIS polarimeter is a dual-beam instrument of the type first
described by \cite{m88} and further discussed by \cite{g91} and
\cite{g95}.  Our aim in this Appendix is to discuss the particular
algorithm we use for reductions, and to demonstrate the robustness of
the system. We use the Stokes Parameters I,Q and U and assume that
the circular polarization parameter V is zero. 

   The polarimeter lies directly under the slit mask of the LRIS.  For
imaging the mask is removed and a mirror is substituted for the
grating.  The beam goes successively through a calibration wheel, a
rotatable super-achromatic half-wave plate, and a calcite
beamsplitter.  The beamsplitter produces parallel
orthogonally-polarized beams separated by $73''$ that go to the camera
and make two images (spectra) on the CCD. The field of view is about
35$''$ square.

  The fundamental assumption we make in analyzing the images is that
the CCD counts are proportional to the linear Stokes parameters
$\rm{I_x}$ and $\rm{I_y}$ at the top of the atmosphere, where $\rm{I_x}
\equiv (I+Q)/2, \rm{I_y} \equiv (I-Q)/2$, and x and y are respectively
the polarization directions of the ordinary and extraordinary (o and e)
rays emerging from the beamsplitter.  We then assume that the
proportionality constants can be factored as follows. When the fast
axis of the waveplate is set to $0\arcdeg$ (ie. parallel to the
x-axis) we have

      $$\rm{B_0 = t_{0x}g_{0b}I_x}    \eqno(1a)$$ 
      $$\rm{T_0 = t_{0y}g_{0t}I_y}    \eqno(1b)$$ 

\noindent where B and T are the CCD counts in the images made by the o
and e rays, respectively; t is the transmission coefficient to the 
beamsplitter and includes atmospheric opacity, telescope magnification,
and the effects of seeing and guiding; and g is the overall gain in the
optical path from the beamsplitter to the extracted CCD counts,
including the effects of flat-fielding and sky subtraction. The
subscript 0 stands for the angular orientation of the waveplate, and the
subscripts b and t stand for the o and e rays and for the bottom and top 
spectra seen in the image displayed by VISTA.

   After an exposure is made at $0\arcdeg$ we make another at $45\arcdeg$, giving

      $$\rm{B_{45} = t_{45y}g_{45b}I_y}    \eqno(2a)$$ 
      $$\rm{T_{45} = t_{45x}g_{45t}I_x}    \eqno(2b)$$

\noindent where we have assumed that the waveplate and the beamsplitter
are perfect, so that a $45\arcdeg$ rotation interchanges $\rm{I_x}$ and
$\rm{I_y}$.  We now assume that there is no polarization dependence in
the transmission coefficients: $\rm{t_{0x} = t_{0y}}$ and $\rm{t_{45x}
= t_{45y}}$ , and that the instrumental gains do not change during the
two exposures:  $\rm{g_{0b} = g_{45b}}$ and $\rm{g_{0t} = g_{45t}}$.
The four images or spectra can now be combined to give the following
three independent quantities, in addition to the total intensity $\rm{I
= I_x + I_y}$.

  $$\Omega \equiv \sqrt{B_0T_0 \over B_{45}T_{45}} = {t_0 \over t_{45}}  \eqno(3)$$

  $$G \equiv \sqrt{B_0B_{45} \over T_0T_{45}} = {g_b \over g_t}         \eqno(4)$$
       
$$q \equiv {{I_x-I_y} \over {I_x+I_y}} = {{{\sqrt{B_0T_{45}} - \sqrt{B_{45}T_0}} \over 
{\sqrt{B_0T_{45}} + \sqrt{B_{45}T_0}}}} \eqno(5)$$

\noindent Eq. (5) can also be written in the suggestive forms

$$q = {{B_0 - GT_0} \over {B_0 + GT_0}} = {{B_0 - \Omega B_{45}} \over 
{B_0 + \Omega B_{45}}}      \eqno(6)$$

\noindent The Stokes Parameter $u$ is similarly found, with waveplate
settings at $22.5\arcdeg$ and $67.5\arcdeg$. Formulae 3--6 are the same
as those described by \cite{m88} although the symbols are different;
our $\Omega$ is their $\omega$. \cite{w92} describes a similar
polarimeter and derives our Eq. (5) (his Eq. 7) but he does not
explicitly discuss G or $\Omega$.

    The quantity $\Omega$ is the ratio of the transmission coefficients
in the two exposures and is useful as a check on the observing
conditions. Figure 5 shows $\Omega$ for the standard star Hiltner 102;
Figure 5a was taken in good conditions and Figure 5b through cirrus.  In both
panels there are two curves, for the $q$ and $u$ observations. In the
top panel the curves overlap.  The middle panel is unusual in the wide
separation between the two curves and the strong wavelength dependence,
due to the changing atmosphere.  In all cases $q$ and $u$ are taken
successively and for the middle panel the total elapsed time was
$147^s$.  Figure 6 shows an imaging $\Omega$ for an exceptionally bad
case, when the seeing was changing rapidly. The two exposures which are
combined to make Figure 6 are each $5^s$ in duration and are separated by
$65^s$. In the center $\Omega = 3.3$, which shows that the seeing
became much worse between the two exposures.

   The quantity G is the gain ratio between the two beams and can be
monitored as a check on the system and on the quality of spectral
extraction.  It has been observed to show variations during a night.
Figure 5c shows a typical curve of G.

   The normalized Stokes Parameter $q$ is given in Eq. 5 in terms of
geometric means that explicitly show how the $t$'s and $g$'s cancel
from each term in the numerator and denominator. The system is
self-calibrated $at ~each ~pixel ~in ~an ~image$ or $at ~each
~wavelength ~in ~an ~extracted ~spectrum$. The calculation is robust
against seeing or opacity changes because each point has its own
values of G and $\Omega$. For example, in spite of the poor sky so
evident in Figs 5b and 6, the values of q and u obtained with those
data were clean and were used as a check on the polarimeter
performance.

   When the seeing is changing (Fig 6) the calculated $q$ at each point
involves a peculiar average over two different seeing disks.  It is not
necessary, however, to convolve the two images to the same psf to get
reliable results.  Note that Eq 5 involves small differences between
large numbers, so that some care must be taken with the data. In
particular, G and $\Omega$ should be calculated internally with the $q$
or $u$ data, and should not be taken from other exposures to be used
with Eq. 6.

   The electric vector of the incident light has a position angle
$\theta = 0.5 \rm{arctan} (u/q)$. For a noise-free situation the
fractional linear polarization $p$ is calculated as $p=\sqrt{q^2+u^2}$,
but this gives a biased result when the S/N is low (S/N $< 5$). In such
cases various de-biasing schemes are available (see \cite{s85}).  Our
data on AGN usually show that $\theta$ varies slowly with $\lambda$ and
so a good approximation to $p$ can be obtained by rotating $(q,u)$ by
$-2\Theta$, where $\Theta$ is a smooth approximation to $\theta$. This
gives $(q',u')$ where $q' \approx p$ and $u' \approx 0$.  $|q'|$ is biased
low by approximately $cos2(\Theta - \theta)$. In our experience this
often is preferable to debiasing the square-root formula
(\cite{s78}).

   When the S/N is low the calculated values of $|q|$ and $|u|$ (Eq. 5)
themselves are biased high (\cite{c83}). In this case it is important
to average the flux values to improve the S/N, and not to average the
$(q,u)$ values.

\newpage

\begin{deluxetable}{cllclll}
\footnotesize
\tablecaption{Journal of Observations. \label{tbl-1}}
\tablewidth{450pt}
\tablehead{
\colhead{Epoch}                   & \colhead{UT Date}                  & 
\colhead{Type\tablenotemark{a}}   & \colhead{Exp\tablenotemark{b}}     &
\colhead{$\Delta \lambda$ (\AA)}  & \colhead{Slit\tablenotemark{c}}    & 
\colhead{Comments} 
}

\startdata
1 &28/10/94 &SPOL &4800 &4000--9000 &1 &Clear        \nl 
2 &29/10/94 &SPOL &6400 &4000--9000 &1 &Cloudy       \nl
3 &31/12/94 &SPOL &2880 &3900--8900 &1 &poor seeing  \nl
4a &16/12/95 &IPOL &3600 &$B$ Band & &seeing $\sim 1.1''$ \nl
4b &16/12/95 &IPOL &3600 &$B$ Band & &seeing $\sim 1.0''$ \nl
5a &17/12/95 &SPOL &3600 &3900--8900 &1.5 &seeing $\sim 0.8''$ \nl
5b &17/12/95 &SPOL &3600 &3900--8900 &1.5 &seeing $\sim 1.5''$ \nl
\enddata

\tablenotetext{a}{SPOL=spectropolarimetry, IPOL=imaging polarimetry}
\tablenotetext{b}{Total exposure time in seconds}
\tablenotetext{c}{Slit width in arc seconds}

\end{deluxetable}

\begin{deluxetable}{crrrrrrrl}
\footnotesize
\tablecaption{Observations of PKS 0116+082. \label{tbl-2}}
\tablewidth{450pt}
\tablehead{
\colhead{Epoch}                                  &
\colhead{$p(\%)$\tablenotemark{a}}               &
\colhead{$\theta$ (deg)\tablenotemark{a}}        &
\colhead{$p(\%)$\tablenotemark{b}}               &
\colhead{$\theta$ (deg)\tablenotemark{b}}        &
\colhead{W$_\lambda$}                            &
\colhead{$F_c^\prime$}                           &
\colhead{$p_c$}                                  &        
\colhead{Comments}
}

\startdata
Aug 91 & & & $5.4 \pm 0.6$ &$75 \pm 3$ &  & & &Tadhunter $V'$ \nl
1 &$5.6 \pm 0.3$ &$13.5 \pm 2.0$ &$4.9 \pm 0.2$ &$8.4 \pm 1.4$ 
   &124 &69 &6.0 &   \nl
2 &  &  &$5.2 \pm 0.2$ &$10.5 \pm 2.1$ &118 &73 &6.3 &cloudy  \nl
3 &$5.8 \pm 0.7$ &$28.7 \pm 2.4$ &$4.3 \pm 0.3$ 
    &$30.6 \pm 1.9$ &122 &70 &5.4 & \nl
4a &$16.7 \pm 0.1$ &$94.9 \pm 0.2$ & & & & & &IPOL $B$ Band  \nl
4b &$15.6 \pm 0.2$ &$91.0 \pm 0.3$ & & & & & &IPOL $B$ Band \nl
5a &$8.5 \pm 0.3$ &$102.0 \pm 1.2$ &$7.3 \pm 0.2$ 
    &$102.1 \pm 0.8$ &99 &89 &8.8 & \nl
5b &$8.9 \pm 0.2$ &$100.9 \pm 2.0$ &$7.0 \pm 0.4$ 
    &$103.0 \pm 1.6$ &99 &89 &8.5 & Seeing 1--2$''$  \nl
\enddata

\tablenotetext{a}{$B$ Band}
\tablenotetext{b}{$V$ Band}

\end{deluxetable}

\begin{deluxetable}{lllll}
\tablecaption{Positions of PKS 0116+082. \label{tbl-3}}
\tablewidth{450pt}
\tablehead{
\colhead{}
&\colhead{RA (J2000)}
&\colhead{$\pm$}
&\colhead{Dec (J2000)}
&\colhead{$\pm$}
}

\startdata
Radio   &$01^h19^m01^s.275$ &$0^s.003$ &$+08\arcdeg29'54''.69$ &$0''.035$ \nl
Optical &$01^h19^m01^s.28$  &$0^s.03$  &$+08\arcdeg29'54''.7$  &$0''.5$ \nl
\enddata

\end{deluxetable}

\clearpage

\newpage

\begin{figure}[t]
 \figurenum{1}
 \plotfiddle{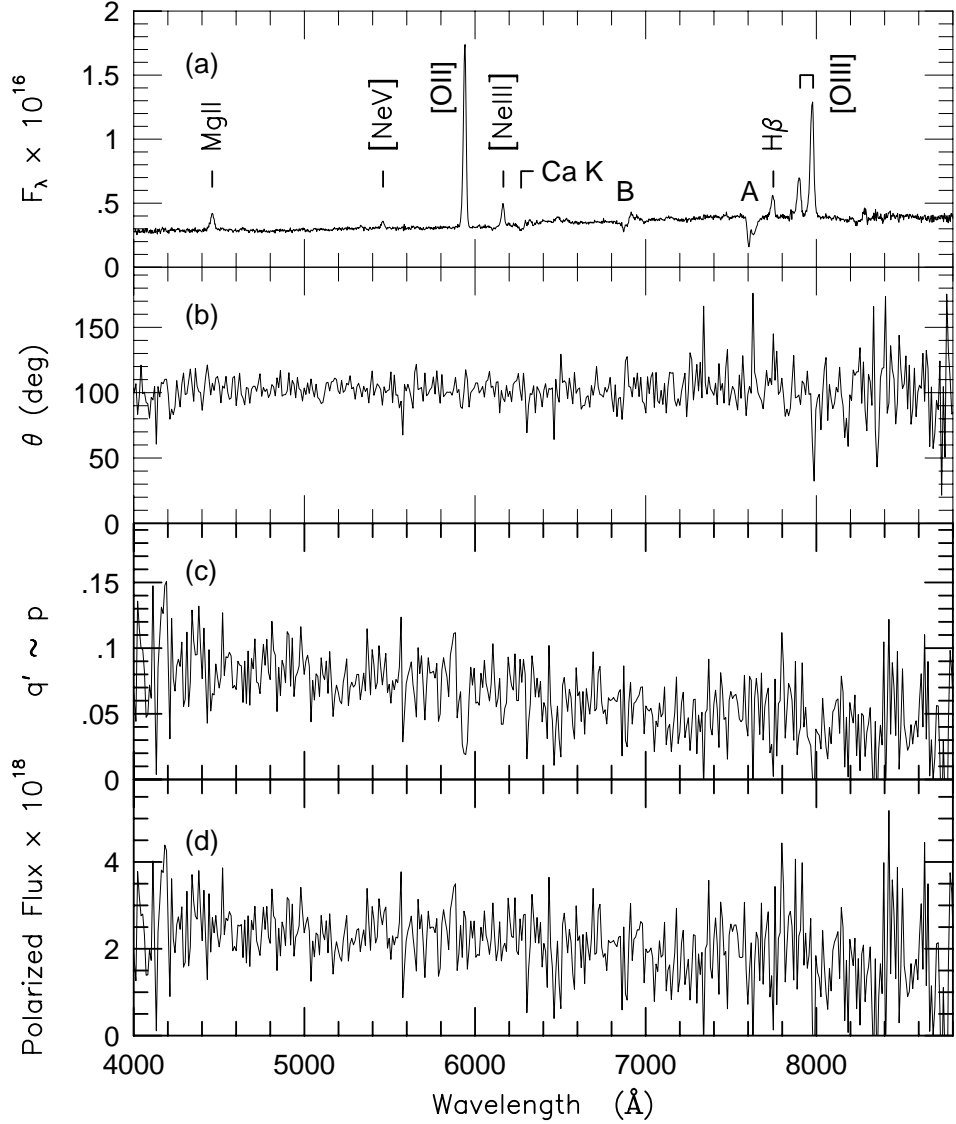}{6.5in}{0}{70}{70}{-216}{-52}
 \caption[fig1.ps]{PKS 0116+082, 1995 December 17. (a) Total Flux.
  The atmospheric absorption bands (A, B, water vapor) have not been
  removed.  (b) Position Angle $\theta$ (c) Fractional Linear
  Polarization, calculated as the ``rotated Stokes parameter" (d)
  Polarized Flux = product of (a) and (c). The Total Flux (a) is shown at
  the original dispersion, 2.49 \AA\ per pixel. Frames (b), (c) and (d)
  are calculated from fluxes which were binned by 4 pixels.
  Note the high continuum polarization in the blue, and the lack of
  polarization in the emission lines. \label{fig1}}
\end{figure}

\begin{figure}[t]
 \figurenum{2}
 \plotfiddle{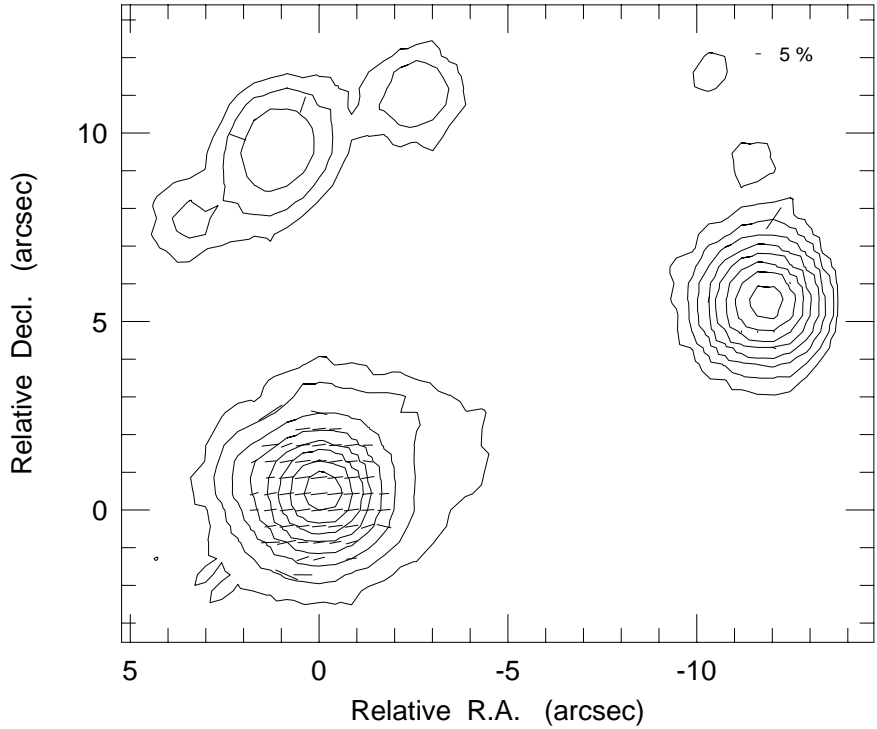}{2.5in}{0}{70}{70}{-216}{-180}
 \caption[fig2.ps]{Polarization image for PKS 0116+082, 1995 December 16.
 Contour levels are 0.5,1,2,4,8,16,32,64\% of the peak. The
 pixels are 0.43$''$ square, and only those vectors with $\rm{S/N} > 2$
 are plotted.  The peak polarization vector near the flux peak
 has $p = 17.3\% \pm 0.2\%$ at $\theta = 96.7\arcdeg \pm 0.4\arcdeg$.
 \label{fig2}}
\end{figure}

\begin{figure}[t]
 \figurenum{3a}
 \plotfiddle{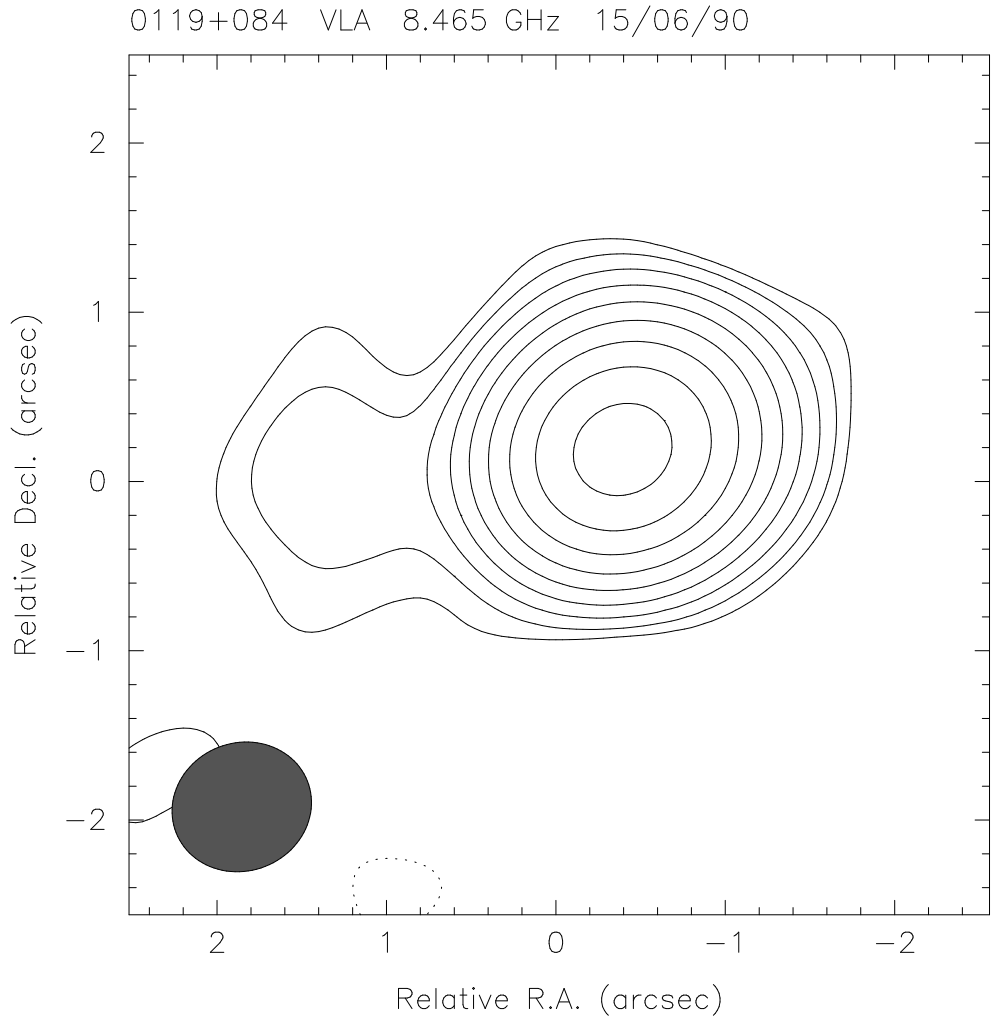}{5in}{0}{70}{70}{-216}{-72}
\caption[fig3.ps]{ VLA images of PKS 0116+082 at 8.5 GHz, observed
 on 1990 June 15. a) A tapered image, with a restoring beam of 0.83 by
 0.76 arcsec in PA $-69\arcdeg$, emphasizing the extended emission.
 The contour levels increase in factors of 2, starting at 2.25 mJy/beam
 (3 times the r.m.s.\ noise in the image), with a peak of 794 mJy/beam.
 \label{fig3a}}
 \end{figure}

\setcounter{figure}{2}

\begin{figure}[t]
 \figurenum{3b}
 \plotfiddle{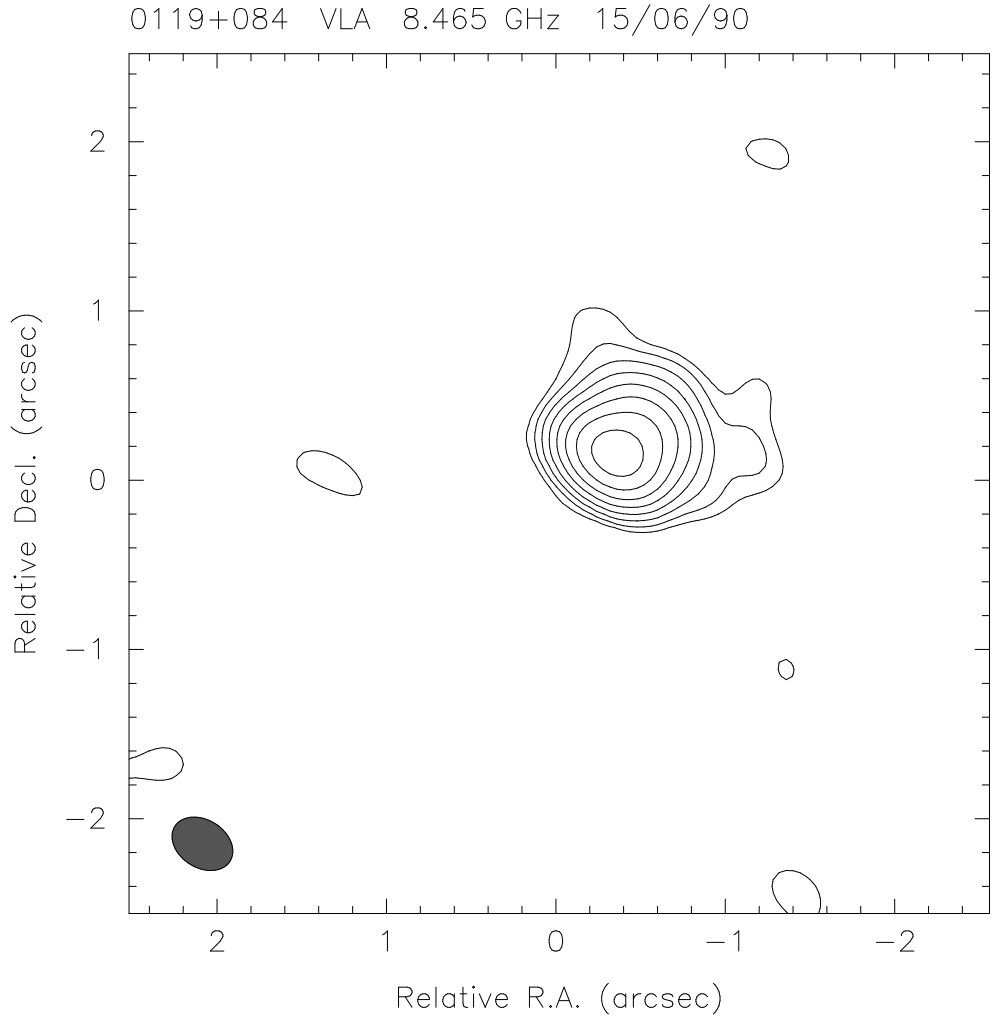}{5in}{0}{70}{70}{-216}{-72}
 \caption[fig3.ps]{ VLA images of PKS 0116+082 at 8.5 GHz, observed
 on 1990 June 15. b) Uniformly weighted image, giving maximum resolution, 
 with a restoring beam of 0.38 by 0.28 arcsec in PA $57\arcdeg$, and contour
 levels starting at 3.26 mJy/beam and a peak of 630 mJy/beam.
 \label{fig3b}}
 \end{figure}

\begin{figure}[t]
 \figurenum{4a}
 \plotfiddle{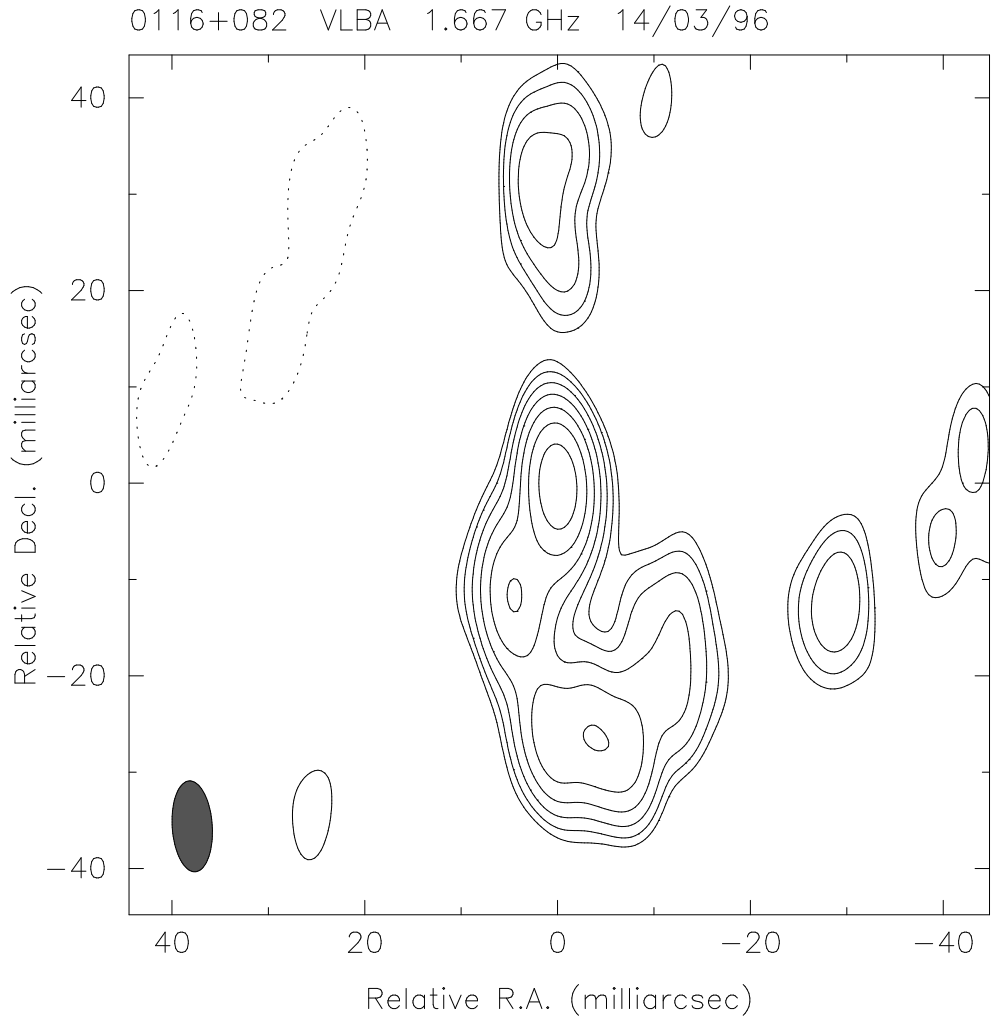}{5in}{0}{70}{70}{-216}{-72}
 \caption[fig4.ps]{ Naturally weighted VLBA images of PKS 0116+082.
 a) at 1.7 GHz, observed on 1996 March 14, with a restoring beam of 9.5
 by 4.2 milliarcsec in PA $4\arcdeg$. The contour levels increase in
 factors of 2, starting at 3.0 mJy/beam (3 times the r.m.s.\ noise in
 the image), with a peak of 362 mJy/beam. \label{fig4a}}
\end{figure}

\setcounter{figure}{3}

\begin{figure}[t]
 \figurenum{4b}
 \plotfiddle{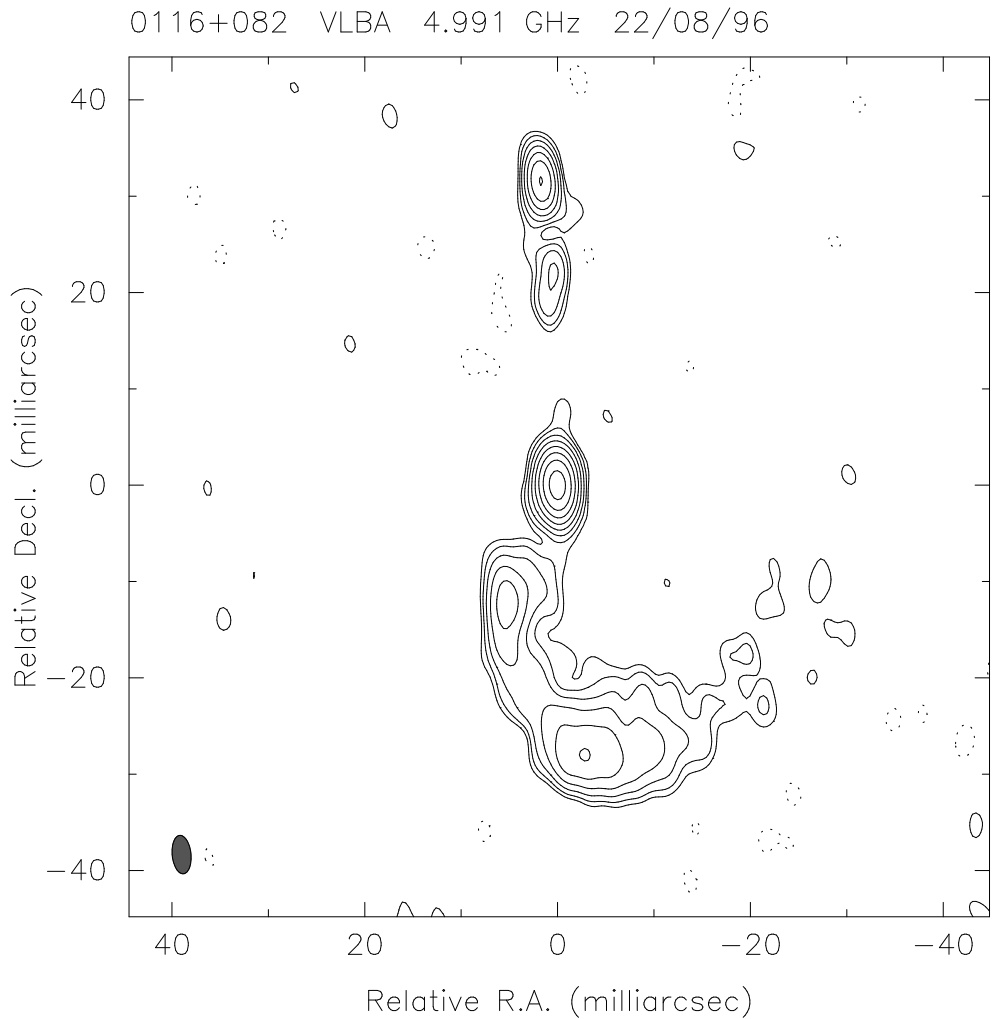}{5in}{0}{70}{70}{-216}{-72}
 \caption[fig4.ps]{ Naturally weighted VLBA images of PKS 0116+082.
 b) at 5.0 GHz, observed on 1996 August 22 with a restoring beam of 
 4.0 by 2.0 milliarcsec in PA $6\arcdeg$. The contour levels increase 
 in factors of 2, starting at 1.0 mJy/beam (3 times the r.m.s.\ noise 
 in the image), with a peak of 183 mJy/beam. \label{fig4b}}
\end{figure}

\begin{figure}[t]
 \figurenum{5}
 \plotfiddle{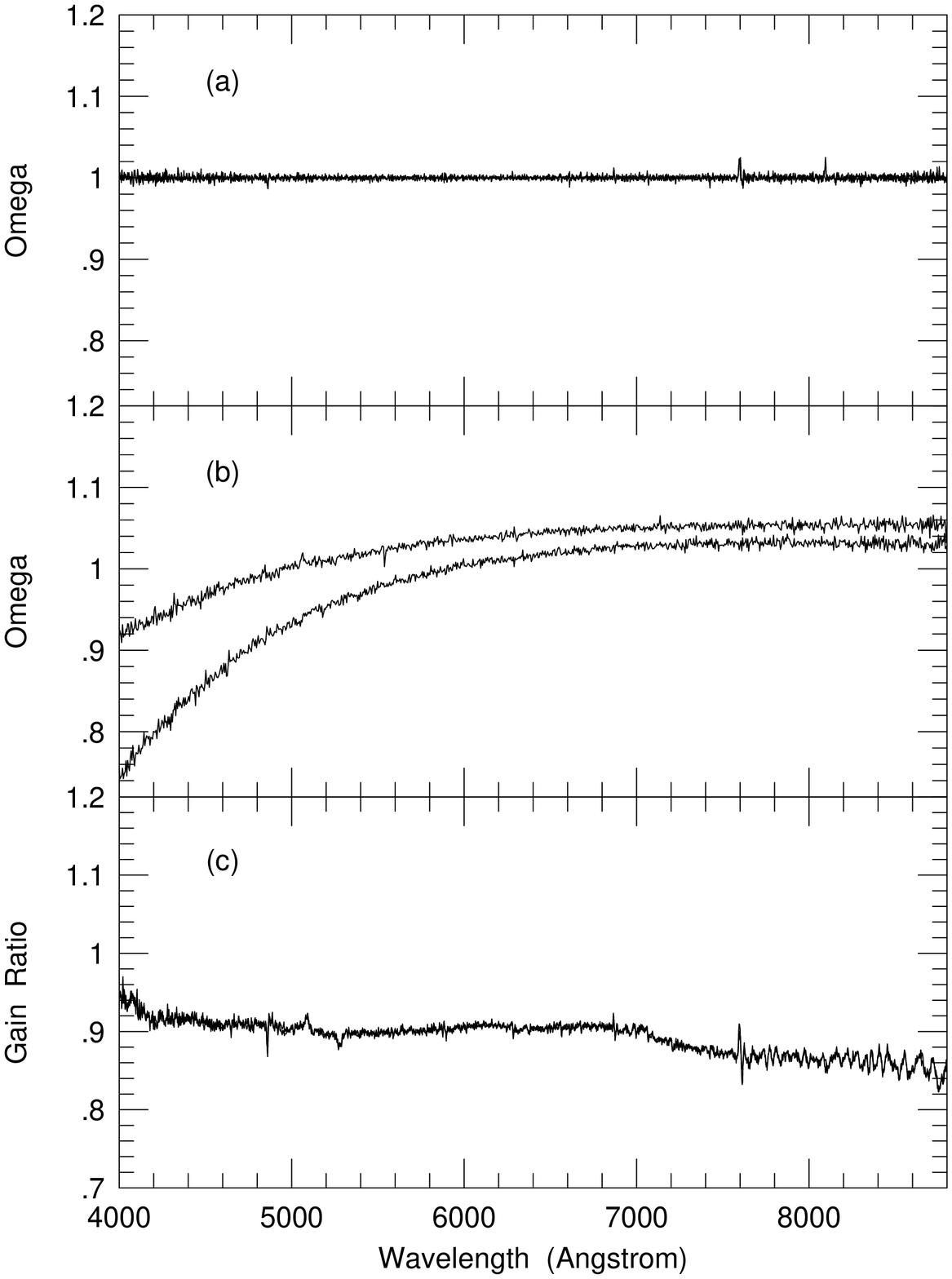}{6.5in}{0}{70}{70}{-216}{-25}
 \caption[fig5.ps]{(a) $\Omega$ (Eq. 3) for Hiltner 102, 
  1994 August, four $20^s$ exposures taken in clear skies (b) $\Omega$ for
  Hiltner 102, 1994 October, four $15^s$ exposures taken through cirrus
  (c) G (Eq. 4) for Hiltner 102, 1994 August, four $20^s$ exposures.
  \label{fig5}}
\end{figure}

\begin{figure}[t]
 \figurenum{6}
 \plotfiddle{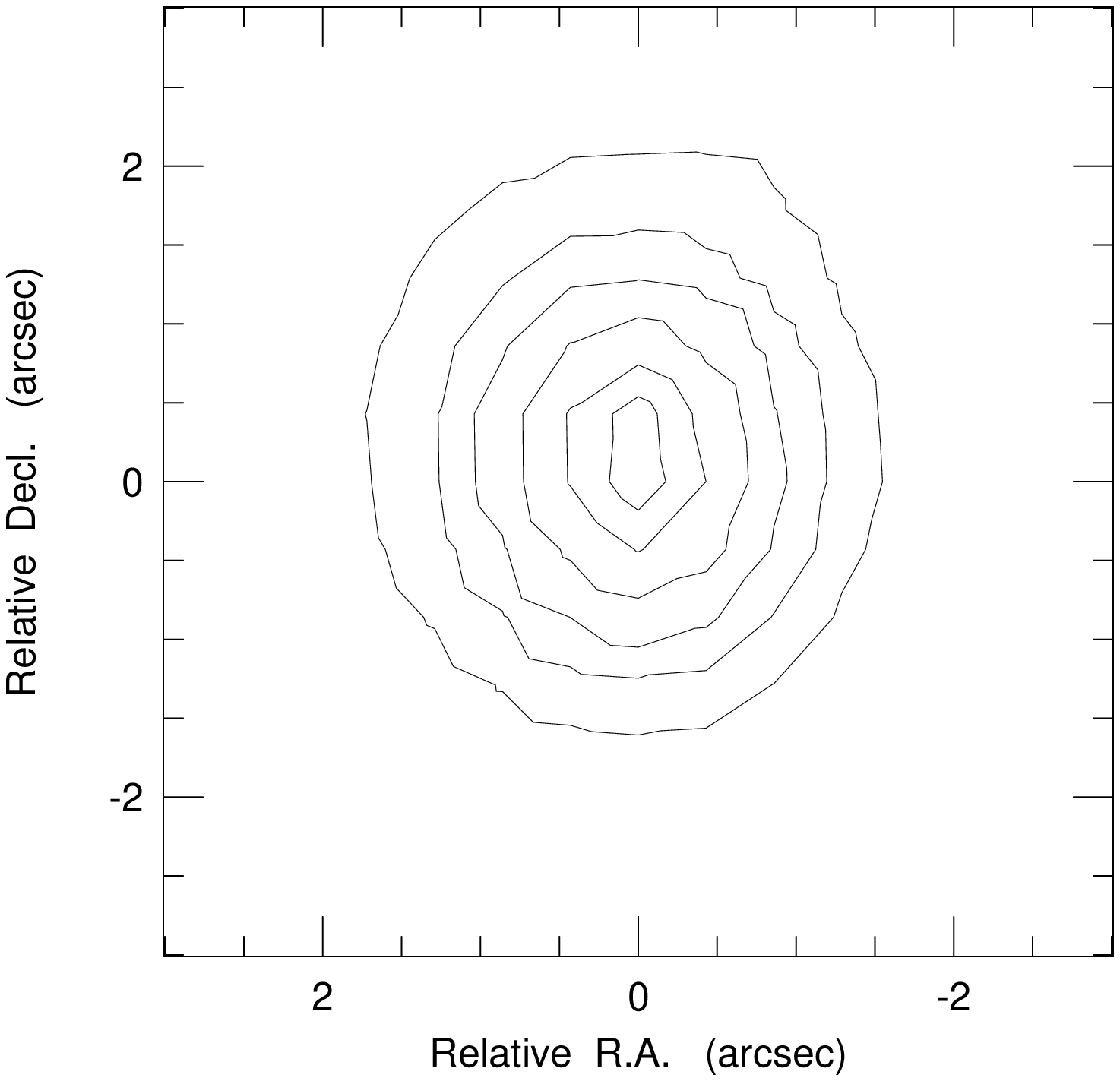}{2.5in}{0}{70}{70}{-216}{-72}
 \caption[fig6.ps]{$\Omega$ (Eq. 3) for star 139 in the field Landolt
 95, 1995 December.  The calculation uses two $5^s$ exposures separated
 by $75^s$. Contours 0.3, 0.5, 0.8, 1.5, 2.5, 3.3. The effect of a
 strong seeing change between the two exposures can be seen.
 \label{fig6}}
\end{figure}

\end{document}